\documentclass[12pt]{iopart}
\usepackage{graphics}
\usepackage{graphicx}
\begin{document}

\newcommand{\refeq}[1]{Eq.\ (\ref{#1})}  
\newcommand{\reffig}[1]{Fig. \ref{#1}}   
\newcommand{\refsec}[1]{Sec.\ \ref{#1}}  

\newcommand{\ket}[1]{\hbox{$\mid \! {#1} \rangle$}}
\newcommand{\bra}[1]{\langle #1 |}
\newcommand{\rket}[1]{\| #1 \rangle}
\newcommand{\rbra}[1]{\langle #1 \|}
\newcommand{\braket}[2]{\langle #1 | #2 \rangle}
\newcommand{\bigbraket}[3]{\langle \, #1 \, | \, #2 \, | \, #3 \, \rangle}
\newcommand{\rbraket}[2]{\langle #1 \| #2 \rangle}
\newcommand{\rbigbraket}[3]{\langle \, #1 \, \| \, #2 \, \| \, #3 \, \rangle}
\newcommand{\com}[2]{{\big[} #1 \, , \, #2 {\big]}}
\newcommand{\anticom}[2]{{\big\{} #1 \, , \, #2 {\big\}}}

\newcommand{\qcg}[6]{ \mbox{$C {}^{#1}_{#4} {}^{#2}_{#5} {}^{#3}_{#6}$} }
\newcommand{\qracah}[6]{ W \left( {#1},{#2},{#3},{#4} ; {#5},{#6} \right) }
\newcommand{\qsixj}[6]{\mbox{$\left\{ \begin{array}{ccc} \!{#1}\! &
                \!{#2}\! & \!{#3}\! \\ \!{#4}\! & \!{#5}\! &
                \!{#6}\! \end{array} \right\}$}}
\newcommand{\qninej}[9]{\mbox{$\left\{ \begin{array}{ccc} \!{#1}\! &
                                \!{#2}\! & \!{#3}\! \\ \!{#4}\! & \!{#5}\! &
                                \!{#6}\! \\ \!{#7}\! & \!{#8}\! & \!{#9}\! 
                                \end{array} \right\}$}}

\newcommand{\qninejsq}[9]{\mbox{$\left[ \begin{array}{ccc} \!{#1}\! &
                                \!{#2}\! & \!{#3}\! \\ \!{#4}\! & \!{#5}\! &
                                \!{#6}\! \\ \!{#7}\! & \!{#8}\! & \!{#9}\! 
                                \end{array} \right]$}}

\newcommand{\vektor}[1]{\mbox{\boldmath $#1$}}
\newcommand{\itensor}[2]{\hbox{$\vektor{#1}^{[#2]}$}}
\newcommand{\itensorcomp}[3]{\hbox{${#1}^{[#2]}_{#3}$}}
\newcommand{\itensorcoupling}[3]{
\hbox{${\big[} #1 \mathbf{\times} #2 {\big]}$}^{[#3]}}
\newcommand{\itensoroutercoupling}[3]{
\hbox{${\big[} #1 \mathbf{\otimes} #2 {\big]}$}^{[#3]}}

\newcommand{\itensordag}[2]{\hbox{$\vektor{#1}^{\dagger[#2]}$}}

\newcommand{\ie}{\mbox{\it i.e. \/}}

\newcommand{\book}[3]{#1: \textit{#2}, #3.}

\title{From density-matrix renormalization group to matrix product states}
\author{Ian P McCulloch}
\address{Institut f\"ur Theoretische Physik C, 
RWTH-Aachen, D-52056 Aachen, Germany}
\ead{ianmcc@physik.rwth-aachen.de}
\date{\today}

\begin{abstract}
In this paper we give an introduction to
the numerical density matrix renormalization group (DMRG) algorithm,
from the perspective of the more general matrix product state (MPS)
formulation. We cover in detail the differences between the
original DMRG formulation and the MPS approach, 
demonstrating the additional flexibility that arises from
constructing both the wavefunction and the Hamiltonian in
MPS form. We also show how to make use of global symmetries, for 
both the Abelian and non-Abelian cases.
\end{abstract}

\maketitle

\section{Introduction}

The DMRG algorithm was introduced by Steven White \cite{White}, as
an algorithm for calculating ground state properties of
principally one-dimensional strongly correlated systems in condensed
matter physics. The connection between DMRG and
matrix product states \cite{Klumper,Fannes,Klumper2,Derria}
(also known as finitely correlated states)
was first made by Rommer and \"Ostlund \cite{rommer}, who identified
the thermodynamic limit of DMRG with a position-independent
matrix product wavefunction. Although DMRG had already
proven itself to be useful empirically, this was an important step in
rigorously establishing the physical basis of the algorithm
due to the concrete and easy to manipulate form of matrix product states.
Work on the spectra of density matrices \cite{Pechel}, later formulated
as scaling of the von Neumann entropy \cite{Entropy} has placed
the algorithm on a firm footing, showing that the required computational
effort (realized via the basis dimension $m$) is essentially a function
of the entanglement of the wavefunction \cite{SchuchEntropy}, 
which for one-dimensional
ground-states scales at worst logarithmically with the system 
size \cite{LogScaling}.

Computationally, MPS algorithms came to the fore with the assistance
of a quantum information perspective, leading to algorithms
for periodic boundary conditions \cite{VerstraetePBC}, and
finite temperature algorithms based on density
operators \cite{Zwolak,VerstraeteDensityOperator,WhiteDensityOperator}.
At around the same time, methods for simulation of real time
evolution were developed in DMRG \cite{UliTime,WhiteTime}, which
can also benefit from MPS formulations \cite{RipollTime}.

The common theme of MPS approaches is to allow algorithms that
operate on multiple, distinct wavefunctions at the same time.
This is possible in the original formulation of DMRG only by
constructing a mixed effective Hilbert space that is weighted
appropriately to represent all of the relevant states simultaneously.
This is inefficient, as the algorithms typically scale as
$O(m^3)$ (or up to $O(m^5)$ for periodic boundary 
conditions \cite{VerstraetePBC}) in the number of basis states $m$,
so increasing $m$ so as to represent multiple states in the same basis 
is typically much slower than performing separate operations
on each basis. In addition, the mixed basis approach lacks flexibility.
While traditional
DMRG programs calculate the wavefunction and
a few (often predetermined) expectation values or 
correlation functions, if instead the wavefunction 
is calculated in the MPS representation
of \refeq{eq:MPWavefunction} it can be saved for later use as
an \textit{input} for many purposes. Perhaps the simplest
such operation beyond the scope of traditional DMRG is to
calculate the \textit{fidelity}, or \textit{overlap}
between the wavefunctions obtained from
separate calculations. In the MPS formulation, this
calculation is straightforward. Nevertheless the determination of
the scaling function
for the fidelity of finite-size wavefunctions for different
interaction strengths, provides a new tool for investigating
phase transitions and crossover phenomena  \cite{Zanardi,QuantumPhaseMPS,Huan}.
Indeed, due to the simplicity of the calculation the
fidelity is likely in the coming years to be the first choice for 
quantitatively determining critical points. Similar measures
of entanglement, such as the concurrence
and single- and two-site entropy \cite{Nielsen,Legeza}, 
are also straightforward
to calculate, hence the MPS formalism allows us to apply
directly the emerging tools of quantum information to the
study of realistic systems in condensed matter physics.
An alternative measure, the Loschmidt Echo \cite{QuanLoschmidt}
is important because, unlike many of the quantum information 
theoretic measures, this is directly accessible in experiments
while showing the rich behavior of the simpler 
measures. The Loschmidt Echo is more
time-consuming to measure numerically as it requires a full
time evolution simulation rather than a direct measurement,
nevertheless it is well within the current state of the 
art \cite{AndreasThesis}.

In this paper, we focus on the case of open boundary condition
matrix product states. This does not preclude calculation of periodic
systems, however the entanglement of such periodic states is
increased such that in the large $L$ limit (where $L$ is the lattice size),
the number of states kept tends to the square of that required for
open boundary conditions \cite{LegezaBoundary}. Algorithms exist for
periodic boundary conditions \cite{VerstraetePBC} and infinite 
systems \cite{VidalThermo} (not to be confused with the
`infinite-size' DMRG algorithm), and the basic formulas introduced
here carry over to these cases, but we do not describe 
the specific algorithms here. In \refsec{sec:MPS}, we introduce
the basic formulation of matrix product states, and formulas
for the fidelity. \refsec{sec:Operators} is devoted to a new
approach, whereby we construct the Hamiltonian operator itself
as an MPS, with many advantages. We cover some remaining
details of the DMRG algorithm in \refsec{sec:DMRG}, before
discussing in detail the use of Abelian and non-Abelian quantum
numbers in \refsec{sec:QuantumNumbers}. We finish with
a few concluding remarks in \refsec{sec:Conclusions},
including some observations on finite temperature states.

\section{Matrix Product States}
\label{sec:MPS}

We denote an MPS on an $L$-site lattice by the form
\begin{equation}
\Tr \sum_{\{s_i\}} A^{s_1} A^{s_2} \cdots A^{s_L} 
\quad \ket{s_1} \otimes \ket{s_2} \otimes \cdots \otimes \ket{s_L} \; ,
\label{eq:MPWavefunction}
\end{equation}
The local index $s_i$ represents an element of a local Hilbert space at site $i$.
The two important cases
we cover here are when $s_i$ runs over a $d$-dimensional local basis for a 
wavefunction $\ket{s_i}$, in which case we refer to this as a matrix product 
wavefunction (MPW),
or $s_i$ is a $d^2$-dimensional local basis for all operators acting on a local site,
which we refer to as a matrix product operator (MPO).
In this paper, we use MPS for a generic state irrespective of the form of the 
local space,
and use MPW or MPO as necessary when the distinction between wavefunctions and 
operators is important.
In general, the matrix product form can also represent
periodic \cite{VerstraetePBC} and infinite (non-periodic) 
states \cite{VidalThermo}, but here we use only the open-boundary form
equivalent to the wavefunction obtained by DMRG \cite{rommer}. 
To enforce this boundary condition, we require the left-most matrix
$A^{s_1}$ to be $1 \times m$ dimensional, and the
right-most matrix $A^{s_L}$ to be $m \times 1$. Here we have introduced
$m$ as the basis size, or dimension of the \textit{matrix basis}
of the $A$-matrices. This quantity is often denoted $D$, or sometimes
$\chi$, in the quantum information literature, but we emphasize
it is exactly the same quantity in all cases.
In general $m$ is position dependent, as we do not
require the $A$-matrices to be square even away from the boundary. 
Because of
the 1-dimensional basis at the boundaries we can regard
the MPS wavefunction to be a sequence of operators attached to
left and right (or outgoing and incoming) vacuum states. This makes
the operator product in \refeq{eq:MPWavefunction} an ordinary number,
so the trace operation can be dropped.

\subsection{Orthogonality constraints}

In practice, a MPS state with no particular constraints on the
form of the $A$-matrices is numerically difficult to handle. 
We are always free to insert some product of a non-singular
$m \times m$ operator $X$ and its inverse $X^{-1}$ in the middle
of our MPS, thus we can apply an arbitrary transformation
to the matrix basis of an $A$-matrix, as long as we make the
corresponding transformation to its neighbor. Using this freedom,
we can transform the $A$-matrices into a form where they
are orthonormalized, that is, we prefer that they satisfy one of two
possible constraints,
\begin{equation}
\begin{array}{rcll}
\sum_s A^s A^{s\dagger} & =&  \textbf{1}  \quad  & \mbox{(right-handed)} \\
\sum_s A^{s\dagger} A^s & =&  \textbf{1}  \quad  & \mbox{(left-handed)}
\end{array}
\label{eq:NormalizationConstraint}
\end{equation}
States satisfying these conditions are orthonormalized in the sense
that if all $A$-matrices to the left of some matrix $A^{s_n}$ are
orthogonalized in the left-handed sense, then the basis on the
left-hand side of $A^{s_n}$ is orthonormal (\textit{ie} the identity operator in
the effective Hilbert space is trivial). Conversely, if all
$A$-matrices to the right of $A^{s_n}$ are orthogonalized in the right-hand
sense, then the basis on the right-hand side of $A^{s_n}$ is orthogonal.
Usually, we want both these conditions to be true simultaneously.
Note that it is not, in general, possible for \textit{all} of the $A$-matrices
(including $A^{s_n}$ itself) to be in orthonormal form at the same time.
There are several ways of transforming an arbitrary MPS into this
normalized form. Two ways that we consider here are the singular
value decomposition (SVD), and the related reduced density matrix,
as used in DMRG \cite{White}. The simplest, and in principle the fastest,
is the SVD, well-known from linear algebra \cite{Higham}. For example,
for the left-handed orthogonality constraint on $A^s_{ij}$,
where we have re-inserted the matrix indices $i,j$, we consider
$s,i$ to be a single index of dimension $dm$, giving an ordinary
$dm \times m$ dimensional matrix, and carry out the singular value
decomposition,
\begin{equation}
A^s_{ij} = \sum_{kl} U^{}_{(si)k} D^{}_{kl} (V^\dagger)_{lj} \; ,
\end{equation}
where $U$ is column-orthogonal, $U^\dagger U = 1$, and $V^\dagger$ is row-orthogonal,
$V^\dagger V = 1$. $D$ is a non-negative diagonal matrix containing
the singular values. This form coincides with the Schmidt decomposition,
where $D$ gives the coefficients of the wavefunction in the Schmidt 
basis \cite{Nielsen}. The matrix $U$ therefore satisfies the left-handed
orthogonality constraint, so we use this as the updated $A$-matrix,
and multiply the $A$-matrix on the right by $DV^\dagger$. This implies
that the $A$-matrix on the right is no longer orthonormalized
(even if it was originally), but we can apply this procedure iteratively,
to shift the non-orthogonal $A$-matrix to the boundary -- or even
beyond it -- at which point the $1 \times 1$ $A$-matrix coincides with
the norm of the wavefunction. An important point here is that we 
can choose to discard some of the states, typically those that
have the smallest singular value. This reduces the matrix dimension
$m$, at the expense of introducing an approximation to our wavefunction,
such that the squared norm of the difference of our approximate
and exact wavefunctions is equal to the sum of the squares of the discarded singular
values. Note however that the singular values only correspond
to the coefficients of the Schmidt decomposition if all of the
remaining $A$-matrices are orthogonalized according to 
\refeq{eq:NormalizationConstraint}. If this is not the case, the singular
values are not useful for determining which states can be safely discarded.

Alternatively, we can construct the reduced density matrix, obtained
by tracing over half of the system. This is achieved by
\begin{equation}
\rho^{s's}_{ij} = \sum_{k} A^{s'*}_{ik} A^s_{jk} \;,
\end{equation}
which is a $dm \times dm$ matrix, with $m$ eigenvalues coinciding with
the values on the diagonal of $D^2$, and the remaining eigenvalues are zero. 
Again, the eigenvalues are only meaningful if the remaining $A$-matrices
are appropriately orthogonalized.
The utility of the density matrix approach over the SVD, is that
we can introduce mixing terms into the density matrix which
can have the effect of stabilizing the algorithm and accelerating
the convergence, which is further discussed in \refsec{sec:DMRG}.

The overlap of two MPS is an operation that appears in
many contexts. For wavefunctions this gives the fidelity 
of the two states, and for operators this is equivalent
to the operator inner product $\braket{A}{B} = \Tr A^\dagger B$
which induces the Frobenius norm. 
Direct expansion of the MPS form yields,
\begin{equation}
    \begin{array}{rcl}
      \braket{A}{B} &=& \displaystyle \sum_{\{s_i\}} 
      \left( \: \Tr \; A^{s_1*} A^{s_2*} \ldots \: \right)
      \left( \: \Tr \; B^{s_1} B^{s_2} \ldots \: \right) \\
      & = & \displaystyle  \Tr \; \sum_{\{s_i\}} 
(A^{s_1*} \otimes B^{s_1})  (A^{s_2*} \otimes B^{s_2}) \ldots \; .
    \end{array}
\label{eq:Overlap}
\end{equation}
Due to the open boundary
conditions, the direct product $E_1 = A^{s_1*} \otimes B^{s_1}$ 
reduces to an ordinary $m \times m$
matrix, after which can construct successive terms recursively, via
\begin{equation}
E_n = \sum_{s_n} A_n^{s_n\dagger} \; E_{n-1} \; B_n^{s_n} \; ,
\label{eq:EMatrix}
\end{equation}
with an analogous formula if we wish to start at the right hand side of the
system and iterate towards the left boundary,
\begin{equation}
F_n = \sum_{s_n} A_n^{s_n} \; F_{n+1} \; B_n^{s_n\dagger} \; .
\end{equation}
For the purposes of numerical stability, it is advisable for
the $A$- and $B$-matrices to be orthogonalized in the same
pattern, that is, $E$-matrices are associated exclusively with the left-hand
orthogonality constraint and $F$-matrices are associated 
with the right-hand orthogonality constraint. If we iterate all the way
to the boundary, the $E$- (or $F$-) matrix ends up as a $1 \times 1$ matrix
that contains the final value of the fidelity. Alternatively, we can iterate from
both ends towards the middle and calculate the fidelity as $\Tr E F^\dagger$.

\subsection{Local updates}

The key notion of the matrix product scheme is that of
\textit{local updates}; that is, we modify,
typically though some iterative optimization scheme, one (or perhaps
a few) $A$-matrix while keeping the remainder fixed. A useful
and flexible alternative is the \textit{center matrix formulation},
where, instead of modifying an $A$-matrix directly, we introduce
an additional matrix $C$ into the MPS,
\begin{equation}
A^{s_1} A^{s_2} \cdots A^{s_n} \; C \; A^{s_{n+1}} \cdots A^{s_L} \; .
\label{eq:CenterMatrixMPS}
\end{equation}
This allows us to preserve orthogonality of the matrices at all times;
matrices $A^{s_i}$ for $i \leq n$ are normalized always according
to the left-handed constraint, and matrices for $i > n$
are normalized according to the right-handed constraint.
We directly modify only the matrix $C$ which simplifies the local
optimization problem as $C$ is just an ordinary matrix.
To introduce the local degrees of freedom, say for the $\ket{s_n}$ states,
we \textit{expand} the basis for $A^{s_n}$. That is, we replace
the $m \times m$ dimensional matrices $A^{s_n}$
with $m \times dm$ matrices $A^{'s_n}$, given by
\begin{equation}
A^{'s_n}_{ij} = \delta_{id+s_n, j} \; ,
\end{equation}
and introduce the $dm \times m$ dimensional center matrix 
\begin{equation}
C_{jk} = A^{s_n}_{ik} \; ,
\end{equation}
with $j = id+s_n$ running over $dm$ states. This doesn't change the
physical wavefunction, as $A^{'s_n} C = A^{s_n}$.
Similarly, we can expand the basis for the $A$-matrix on the right
side of $C$, to give the effect of modifying either a single
$A$-matrix, or two (or more) at once.
In the center matrix formulation, the singular value decomposition
required for truncating the basis is simply the ordinary SVD
on the matrix $C = U D V^\dagger$, and we multiply (for a right-moving
iteration)
$A^{'s_n} U$, which preserves the left-handed orthogonality constraint,
and $DV^\dagger A^{s_{n+1}}$ which is not orthogonal, but becomes so
when we again expand the basis to construct the new $C$ matrix.
The density matrix in the center matrix formulation is simply
$\rho = C C^\dagger$ or $\rho = C^\dagger C$ for left and
right moving iterations respectively. For readers already familiar
with DMRG, the center matrix corresponds exactly with
the matrix form of the \textit{superblock
  wavefunction} \cite{UliReview}.

\subsection{Matrix product arithmetic}

The utility of the MPS approach is realized immediately upon
attempting manipulations on the wavefunction \refeq{eq:MPWavefunction}.
Suppose we have two distinct MPS, defined over the
same local Hilbert spaces,
\begin{equation}
\begin{array}{rcl}
\ket{\Psi_A} &=& A^{s_1} A^{s_2} \ldots A^{s_L} 
\; \ket{s_1} \ket{s_2} \ldots\\
\ket{\Psi_B} &=& B^{s_1} B^{s_2} \ldots B^{s_L}
\; \ket{s_1} \ket{s_2} \ldots\\
\end{array}
\end{equation}
The superposition $\ket{\psi_C} = \ket{\Psi_A} + \ket{\Psi_B}$ is 
formed by taking
the sum of the matrix products, $A^{s_1}A^{s_2}\ldots + B^{s_1}B^{s_2}\ldots$,
which can be factorized into a new MPS, 
$\ket{\Psi_C} = C^{s_1} C^{s_2} \ldots$, with
\begin{equation}
C^{s_i} = A^{s_i} \oplus B^{s_i} \; .
\label{eq:MPSSum}
\end{equation}
To preserve the one-dimensional basis at the boundaries, the direct sum is
replaced by a concatenation of columns or rows, for the left and
right boundary respectively. This procedure results in an
MPS of increased dimension, $m_C = m_A + m_B$. Thus, after constructing
these matrices we need to re-orthogonalize the state, and then
we can, if necessary, truncate the basis size to a given truncation
error, which is well defined here and 
measures the exact difference between the original
and truncated wavefunctions. Alternatively, the normalized and
truncated MPS $\ket{\Psi_C}$ can be constructed directly, by
calculating the overlap matrices $E$ between $\ket{\Psi_A}$ and
$\ket{\Psi_B}$. From the $E$-matrices introduced in \refeq{eq:EMatrix}, 
we can construct
directly the orthogonalized reduced density matrices of $\ket{\Psi_C}$
and truncate the basis as required, in a single step.
This approach has better computational scaling than
the two-step procedure of first orthogonalizing and then truncating, especially
when the number of MPS in the superposition is large. But in general,
iterative optimization approaches, where we use a DMRG-like algorithm
to optimize the overlap 
$\bra{\Psi_C} \, \left(\ket{\Psi_A} + \ket{\Psi_B}\right)$, have even
better performance scaling with large $m$ or more states in the
superposition.

\section{Operators}
\label{sec:Operators}

A useful generalization of the MPS structure 
\refeq{eq:MPWavefunction} is to use it to represent an operator (an MPO)
instead of a wavefunction. This has been used before for
calculating finite-temperature density 
matrices \cite{VerstraeteDensityOperator},
but here we instead want to use this structure to represent the
Hamiltonian operator itself. 
All Hamiltonian operators with finite-range interactions have an
\textit{exact} MPS representation with a relatively small
matrix dimension $M$. For example, the Ising model in a transverse
field has a dimension $M=3$, and the fermionic Hubbard model has dimension $M=6$.
We use the capital letter to distinguish
from the dimension of the wavefunction, $m$. Similarly, the local dimension
of the upper index is denoted here by $D$, which is usually
just equal to $d^2$, but slightly complicated in the case of
non-Abelian quantum numbers (see \refsec{sec:NonAbelian}).

We denote an MPO by the form
\begin{equation}
\sum_{\{s_i, s'_i\}} M^{s'_1 s^{}_1} M^{s'_2 s^{}_2} \cdots M^{s'_L s^{}_L} 
\quad \ket{s'_1}\bra{s^{}_1} \otimes \ket{s'_2}\bra{s^{}_2} \cdots \; ,
\label{eq:MPOperator}
\end{equation}
where again we require that the first and last dimensions are $M=1$,
for open boundary conditions. 

The orthogonality constraint used previously for the MPS,
\refeq{eq:NormalizationConstraint}, is not appropriate
for Hamiltonian operators. When applied to an operator, the usual
orthogonality constraints utilize the (Frobenius) operator norm,
which scales exponentially with the dimension of the Hilbert space.
With this normalization, components of an MPO Hamiltonian, such as the identity operator
or some interaction term, tend to differ in magnitude by some factor
that increases exponentially with the lattice size. Arithmetic
manipulations on such quantities is a recipe for 
catastrophic cancellation \cite{Higham}
resulting in loss of precision. Mixing operators with
a unitary transformation (for example
$O_1' = O_1\cos \theta + O_2\sin\theta, O_2' = -O_1\sin\theta + O_2\cos\theta$),
will lead to a disaster if $O_1$ and $O_2$ differ by a sufficiently large order of magnitude,
$\sim 10^{16}$ for typical double-precision floating point arithmetic.
But such rotations are inevitable in the orthogonalization procedure
because in general the
operator inner product $\braket{O_1}{O_2} = \Tr O_1^\dagger O_2$ will not be zero.
Instead we completely avoid mixing different rows/columns
of the operator $M$-matrices, only collapsing a row or column if it
is exactly parallel with another row or column. In this case, the
actual norm of each component of the $A$-matrices is irrelevant,
as they are never mixed with each other (but see also
the discussion of the single-site algorithm in \refsec{sec:DMRG}). 
For
physical Hamiltonian operators this remains essentially optimal, 
with the minimum possible matrix dimension $M$. The only operators for which 
this orthogonalization scheme does not produce an optimal representation
are operators that have a form analogous
to an AKLT \cite{AKLT} state where the local basis states of the $S=1$ chain
are replaced by local operators.
The resulting operator contains an exponentially large
number of $N$-body interactions for all 
$N \rightarrow \infty$. We know of no physical
Hamiltonians of this form.

Given a Hamiltonian as a sum of finite-range interactions,
it is possible to construct the operator $M$-matrices such
that they are entirely lower (or upper) triangular matrices, thus
in principle we can `normalize' the matrices via some kind of generalized
$LU$ or $QR$ decomposition. In practice we don't need to do this,
as the Hamiltonian can easily be constructed in lower-triangular form
from the beginning.
Imposing, again without loss of generality, that the top-left and
bottom-right components of the operator $M$-matrices are equal to
the identity operator $I$, we can construct the sum of $1$-site
local terms $H = \sum_i X_i$ as a position-independent MPS,
\begin{equation}
M = \left( \begin{array}{cc} I & 0 \\ X & I \end{array} \right) \; ,
\label{eq:WState}
\end{equation}
which we regard as a $2\times 2$ matrix, the elements of which are
$d \times d$ dimensional local operators.
For nearest-neighbor terms, $H = \sum_i X_i Y_{i+1}$, we have
\begin{equation}
M = \left( \begin{array}{ccc} 
I & 0 & 0 \\ 
Y & 0 & 0 \\
0 & X & I \end{array} \right) \; ,
\end{equation}
with the obvious generalization to $N$-body terms.
The direct sum and direct product of lower triangular matrices
is itself lower triangular, thus this form can be preserved throughout
all operations. For open boundary conditions, the left (right)
boundary $1 \times M$ (or $M \times 1$) matrices are initialized to
$(0, \ldots, 0, I)$ and $(I, 0, \ldots, 0)^T$ respectively.

The principal advantage of formulating the Hamiltonian operator
(and indeed, \textit{all} operators needed in the calculation) in this
way that it can be manipulated extremely easily, amounting to
a symbolic computation on the operators. 
This is in contrast
to the ad hoc approach used in past DMRG and MPS approaches
where the block transformations required for each operator
are encoded manually, with limited scope for arithmetic operations.
In particular, the
sum of operators is achieved via \refeq{eq:MPSSum}. Products of operators are
achieved by matrix direct product; given MPO's $A$ and $B$,
the product $C = A B$ is given by the matrices
\begin{equation}
C^{s's} = \sum_{t} A^{s't} \otimes B^{ts} \; .
\label{eq:OpProd}
\end{equation}
An implication of this is that the square of an MPO has a matrix dimension
of at most $M^2$, which, since $M$ is usually rather small,
means that it is quite practical to calculate expectation values
for higher-order moments, for example of the \textit{variance}
\begin{equation}
\sigma^2 = \langle{H^2}\rangle - \langle H\rangle^2\ 
= \langle (H-E)^2 \rangle \; ,
\label{eq:Variance}
\end{equation}
which has been mentioned previously \cite{Weichselbaum} as it
gives a rigorous \textit{lower bound} on the energy (although
with no guarantee that this corresponds to the ground-state).
In practice this lower bound is too wide to be useful in all but
the simplest cases, but of more interest is the property
that the variance is, to first order, proportional
to the squared norm of the difference between the exact and
numerical wavefunctions, and therefore also proportional to
the truncation error \cite{Next} (see \refsec{sec:DMRG}). Thus,
this quantity gives a quantitative estimate of the goodness of the
wavefunction even for situations where the truncation error
is not available.
For our numerical MPS algorithms
the variance takes the role of the precision $\epsilon$ in
numerical analysis \cite{Higham} via $\epsilon \sim \sqrt{\sigma^2}$.

Of a similar form to the product of two operators, 
the action of an operator $M$ on a wavefunction $\ket{A}$
gives a wavefunction $\ket{B}$ with matrix elements,
\begin{equation}
B^{s'} = \sum_{s} M^{s's} \otimes A^{s} \; .
\label{eq:OpStateProd}
\end{equation}
The MPO formulation also gives a natural form for the evaluation of
expectation values, similarly to the fidelity of \refeq{eq:Overlap},
\begin{equation}
\bigbraket{A}{M}{B} = \sum_\alpha \Tr E^\alpha_n F^{\alpha\dagger}_n \; ,
\label{eq:Expectation}
\end{equation}
where the $E$- and $F$-matrices now have an additional index $\alpha$
that represents
the terms in the MPO $M$,
with a recursive definition
\begin{equation}
\begin{array}{rcl}
E^{\alpha'}_n &=& \sum_{s',s,\alpha} M^{s's}_{\alpha'\alpha} 
A^{s'_n\dagger} E^{\alpha}_{n-1} B^{s_n} \\
F^{\alpha'}_n &=& \sum_{s',s,\alpha} M^{s's}_{\alpha'\alpha}
A^{s'_n} F^{\alpha}_{n+1} B^{s_n\dagger}
\end{array}
\end{equation}
where again we can either iterate all the way to a boundary, at which point
the $\alpha$ index collapses down to one-dimensional and the
$E^\alpha$ or $F^\alpha$ are $1 \times 1$ dimensional matrices containing
the expectation value, or we can iterate from both boundaries and meet
at the middle, where our expectation value is given by the scalar product
\refeq{eq:Expectation}.

Incidentally, given that the identity operators
occur in a fixed location in the operator $M$-matrix
(ie. at the top-left and bottom-right of the $M$-matrix)
this fixes the index $\alpha$ of the
reduced Hamiltonian and identity operators for the left and right partitions of the system.
That is, in
the application of the Hamiltonian $E^\alpha,F^\alpha$ matrices to the wavefunction
we are guaranteed that the $\alpha = 1$ component
of $E^\alpha \otimes F^{\alpha\dagger}$ corresponds precisely to
$H_L \otimes I_R$, and the $\alpha = M$ component corresponds to
$I_L \otimes H_R$. Thus, even after an arbitrary series of MPO computations
we can still identify exactly which component of the $E,F$ matrices
corresponds to the block Hamiltonian.
This is useful for eigensolver preconditioning
schemes \cite{MyThesis}, for example to change to a basis where the block Hamiltonian
is diagonal.

\begin{figure}[ht]
\centering
\includegraphics[
width=0.8\textwidth,clip]{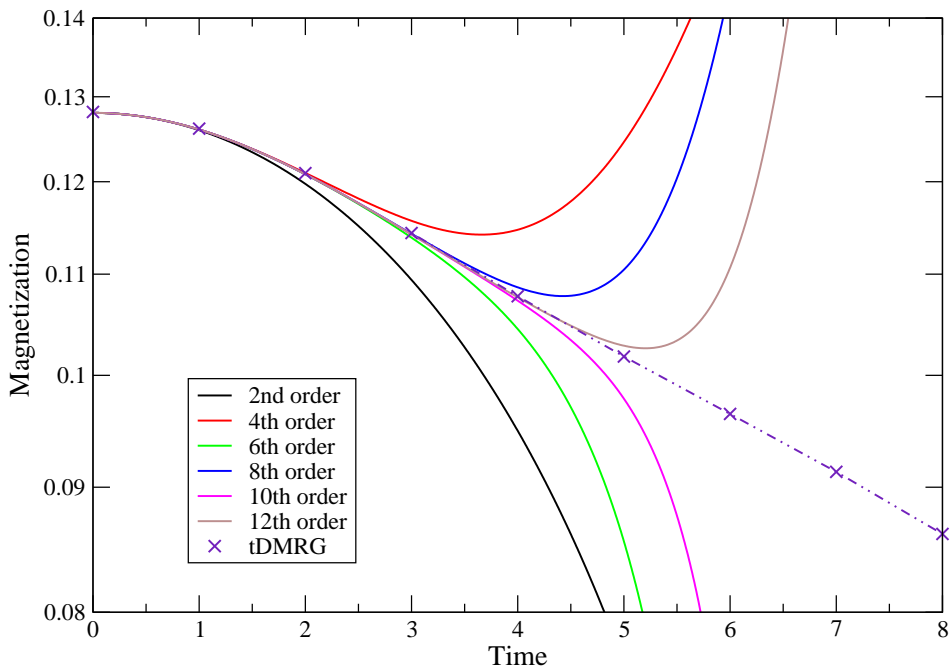}
\caption{Polynomial expansion for the relaxation of
the impurity magnetization in the SIAM, up to order $t^{12}$. The $\times$ symbols
denote the 
impurity magnetization calculated via adaptive time
DMRG, the dashed line is a guide to the eye.
Parameters of the calculation were (in units of bandwidth),
$\Gamma = 0.1, U=0.2, h_0 = 0.1, \epsilon_{d0} = 0.05$, on a
log-discretized Wilson
chain with $\Lambda =1.8$. At time $t=0$, the
Hamiltonian was switched to $h_1=0$ and $\epsilon_{d1} = -0.1$.
}
\label{fig:TimeExample}
\end{figure}

As an example of the utility of this approach,
\reffig{fig:TimeExample} shows the time evolution of the magnetization
of the impurity spin in the single impurity Anderson model (SIAM),
where the ground-state is obtained with a small magnetic field which
is then turned off at time $t=0$. The MPS operator approach readily allows
the evaluation of the commutators required for a small $t$ expansion
of the expectation value of an observable in the Heisenberg picture,
\begin{equation}
A(t) = A + \frac{it}{\hbar}[H,A] - \frac{t^2}{2!\hbar^2}[H,[H,A]]
- \frac{it^3}{3!\hbar^3}[H,[H,[H,A]]] + \cdots \; .
\end{equation}
Since the number of terms in the repeated commutator will, in general,
increase exponentially 
the accessible time-scales from such an expansion are clearly
limited. Nevertheless this is a very fast and accurate way to obtain
short-time-scale dynamics, and in this example $12^{\mathrm th}$ order
is easily enough to determine the $T_1$ relaxation rate.
For this calculation, the terms up to $t^8$
took a few seconds to calculate, while the $t^{10}$ term
took 6 minutes and the $t^{12}$ term took just over an
hour, on a single processor 2GHz Athlon64. This time was divided
between calculating the MPO matrix elements (the dimension of which
was just over 2500 at the impurity site), and calculating the expectation value itself.

\section{DMRG}
\label{sec:DMRG}

We now have all of the ingredients necessary to construct
the DMRG algorithm for determining the ground-state. Indeed,
given the previous formulations, the DMRG itself is rather simple;
using the center matrix formulation, we iteratively improve
the wavefunction locally by using the center matrix $C$ as
input to an eigensolver for the ground-state of the Hamiltonian.
The details of this calculation are precisely as for DMRG,
already covered elsewhere \cite{UliReview}. 

An important component of DMRG, which has been neglected in some
matrix product approaches, is the truncation error.
If only a single site is modified at a time, the maximum number
of non-zero singular values is bounded by the matrix dimension $m$, thus
the matrix dimension cannot be incrementally increased as the calculation progresses.
Some way of avoiding this limitation is practically essential for 
a robust algorithm. The original DMRG formulation
\cite{White} solved this problem by modifying two A-matrices simultaneously,
equivalent to expanding the matrix dimension for both the left and right
$A$-matrices in \refeq{eq:CenterMatrixMPS} so that the center matrix has
dimension $md \times md$. A scheme for single-site algorithms that
avoids the limit on the singular values was introduced by 
Steven White \cite{WhiteSingle}, which uses a mixed density matrix
constructed by a weighted sum of the usual reduced density matrix
and a perturbed density matrix formed by applying the
$E^\alpha$-matrices (on a right-moving sweep) 
or $F^\alpha$-matrices (on a left-moving sweep) of the Hamiltonian,
\begin{equation}
\rho' = \rho + c \sum_\alpha E^\alpha \rho E^{\alpha\dagger} \; ,
\end{equation}
where $c$ is some small factor that fixes the magnitude of the fluctuations.
This solves nicely the problem of the bound on the number of singular
values and introduces fluctuations into the density matrix
that give the algorithm good convergence properties, often better 
than the
two-site algorithm. A minor problem is that the scaling of the
$E^\alpha$ matrices is not well defined, in that we can scale
the $E^\alpha$-matrices by an arbitrary $M \times M$ non-singular matrix 
$X_{\alpha'\alpha}$, 
while at the same time scaling the $F$-matrices by $X^{-1}$.
For one- and two-site terms, there is an `obvious' scaling factor to use,
whereby the scaling factors are chosen such that the (Frobenius) operator
norm of the $E$ and $F$-matrices are identical, but I don't know how
this would apply more generally.
An alternative that appears interesting is to apply the full Hamiltonian
to a density operator for the
full system, $\rho' = \rho + c \Tr_R H (\rho \otimes I) H$, 
constructed from the left (right) reduced
density matrix and the right (left) identity operator, followed
by a trace over the right (left) partition. In MPS form, this
operation is
\begin{equation}
\rho' = \rho + c \sum_{\alpha,\beta} E^\alpha \rho E^{\beta\dagger} 
\; G_{\alpha\beta} \; ,
\end{equation}
where $G_{\alpha\beta} = \Tr F^{\alpha\dagger} F^{\beta}$ 
is an $M \times M$
coefficient matrix. However, this scheme often fails; incorporating the
$G_{\alpha\beta}$ matrix reduces the fluctuations such that
$E^\alpha \rho E^{\beta\dagger} \; G_{\alpha\beta}$ differs little from
$\rho$ itself and the algorithm typically fails to reach the ground-state.
The single-site algorithm \cite{WhiteSingle} corresponds
to choosing $G_{\alpha\beta} = \delta_{\alpha\beta}$.
A useful compromise appears to be using only the \textit{diagonal}
elements, such that 
$G_{\alpha\beta} = \delta_{\alpha\beta} \Tr F^{\alpha\dagger}
F^{\beta}$,
but this is surely not the last word on this approach.
Both the two-site and mixed single-site algorithm inevitably result
in a reduction in the norm of the wavefunction by truncating
the smallest non-zero eigenvalues of the density matrix. The sum of the
discarded eigenvalues, summed over all iterations in one sweep,
is equal to the truncation error $\tau$, familiar
from DMRG \cite{White} (but note that it is common in the literature
to quote an average or maximum truncation error \emph{per site}). 
This quantity is useful in giving an estimate of the 
error in the wavefunction in this is, for a properly converged wavefunction, 
proportional to the norm of the difference between the
exact ground-state and the numerical approximation.
The presence of the truncation error explains why the 
bare single-site algorithm,
despite having slightly better variational wavefunction than the
two-site or mixed single-site algorithms \cite{SierraVariational},
converges much slower; the single site algorithm is a highly constrained
optimization within an $m$-dimensional basis, whereas the
two-site and mixed single-site algorithms are selecting the optimal
$m$ basis states out of a pool of a much larger set of states,
namely the discarded states at each iteration (total $\sim Lm$
states). While the notion of truncation error remains useful in MPS
algorithms, for the purposes of error analysis we much prefer the
variance \refeq{eq:Variance} as being a direct measure of the accuracy
of the wavefunction, independent of the details of a particular
algorithm \cite{Next}.

Low-lying excited states can be constructed using this algorithm.
This has been done in the past in DMRG by targeting multiple eigenstates
in the density matrix, but the MPS formulation allows a substantial
improvement. Namely, it is easy to incorporate into the eigensolver
a constraint that the obtained wavefunction is orthogonal to
an arbitrary set of predetermined MPS's. That is, after
constructing the MPS approximation to the ground-state, we can,
as a separate calculation, determine the first excited state by
running the algorithm again with the constraint that our obtained
wavefunction is orthogonal to the ground-state. This is achieved
by constructing the $E$-matrices that project the set of states to
orthogonalize against onto the local Hilbert space. These
matrices are precisely those used in constructing the fidelity,
\refeq{eq:Overlap}, thus given the center matrix of some state $C_X$,
we project this onto the current Hilbert space $C'_X = E C_X F^\dagger$,
and as part of the eigensolver, orthogonalize our center matrix
against this state. This is a very fast operation, much faster than
even a single Hamiltonian-wavefunction multiplication. So it is
quite practical to orthogonalize against a rather large number of states,
the practical limit is rather on numerical limitations in orthogonalizing
the Krylov subspace in the eigensolver. If this is combined
with an eigensolver capable of converging to eigenvalues in the middle
of the spectrum (say, the lowest eigenvalue larger than some bound $E_0$),
then we need only a small number of states to orthogonalize against,
say half a dozen states immediately above $E_0$ in energy. In our
numerical tests it seems to be rather common to skip eigenvalues,
which is why we cannot simply orthogonalize against a single state. 
With a larger number of states to orthogonalize
against, skipping eigenvalues is less of a problem as we are
likely to recover the missing eigenstate on a later calculation.
Using this approach, quantities such as 
the level spacing statistics can be determined for system sizes
far beyond exact diagonalization \cite{Next}.

\section{Quantum Numbers}
\label{sec:QuantumNumbers}

An important feature of matrix product states is that they
can easily be constrained by quantum numbers representing
the global symmetries of the Hamiltonian, as long as
the symmetry is not broken by the spatial geometry of the MPS.
For example, internal rotational symmetries such as $U(1)$
and $SU(2)$ \cite{NonAbelian} can be maintained exactly, but
for a real-space lattice we cannot utilize the momentum
in the same way because the representation itself violates
this symmetry\footnote{See also Ref.~ \cite{Translation}
for a real-space approach to constructing momentum eigenstates.}.
To achieve this, we impose a symmetry constraint on
the form of the $A$-matrices, so that they are 
\textit{irreducible tensor operators}. That is, under a symmetry
rotation the matrix $A^s$ for each local degree of freedom $s$
transforms according to an 
irreducible representation $D(j_s)$ of the global symmetry group. 
This is a very general procedure,
that is applicable to essentially all MPS approaches 
and generalizations thereof.

\subsection{Abelian symmetries}

For Abelian symmetries, the representations are one-dimensional
therefore the set of quantum numbers labeling the irreducible
representations also forms a group, which we can write as
\begin{equation}
D(j) \otimes D(k) = D(j+k) \; ,
\label{eq:AbelianGroupRep}
\end{equation}
for two representations $D(j)$ and $D(k)$,
where $j+k$ denotes the group operation.
Thus to incorporate Abelian symmetries into the algorithm
we simply attach a quantum number to all of the labels appearing
in the MPS, with the constraint that each $A$-matrix transforms
irreducibly, so that the only non-zero matrix elements are
\begin{equation}
A^k_{q'q} \quad \quad q' = q+k \; ,
\label{eq:QuantumNumbers}
\end{equation}
where $k, q', q$ are the quantum numbers attached to the local basis state
and left and right matrix basis states respectively.
We have suppressed here indices not associated with a quantum number,
a convention which will be followed for the remainder of the paper.

By convention, for our open boundary condition MPS we choose the right
hand vacuum state to have quantum number zero. The symmetry 
constraint \refeq{eq:QuantumNumbers} then implies that the quantum number
at the left hand vacuum will denote how the state as a whole transforms
(the \textit{target state}, in DMRG terminology). This is the
only real difference between DMRG and MPS algorithms, in that
the DMRG convention is to construct both blocks starting from
a scalar (quantum number zero) vacuum state, so that the superblock
wavefunction is a tensor product of two ket states,
\begin{equation}
\ket{\Psi} = \sum_{uv} \psi_{uv} \ket{u} \otimes \ket{v} \; ,\quad \quad
u+v = \mbox{target} \; ,
\end{equation}
whereas for the MPS formulation the superblock wavefunction is represented
by a scalar operator with a tensor product basis
$\ket{u} \otimes \bra{v}$ with quantum numbers $u=v$. This means
that, in contrast to the usual formulation of DMRG, 
the target quantum number is not encoded in the superblock but
rather in the left vacuum state. A consequence
of this is that DMRG is capable of representing simultaneously
states with different quantum numbers, but an MPS is 
not.
This is an important detail, for example in the calculation of dynamical
correlations, as both the correction vector \cite{CorrectionVector} and 
the similar DDMRG \cite{DDMRG} algorithm require a basis optimized
for both the ground-state $\ket{\Psi}$ and the so-called Lanczos-vector
$A \ket{\Psi}$, where $A$ is some operator that may not be scalar.
However, the MPS formulation allows significant 
optimizations to these algorithms
whereby the the calculation of the ground-state is decoupled
from that of the Lanczos vector \cite{Weichselbaum,Andreas} and the two
need never appear in the same basis. 

\subsection{Non-Abelian symmetries}
\label{sec:NonAbelian}

If the symmetry group is large enough that some elements do not 
commute with each other, then it is no longer possible to construct
a basis that simultaneously diagonalizes all of the generators
hence the approach of the previous section needs some modification.
Instead, we label the basis states by quantum numbers that
denote the representation, which is no longer simply related to
the group elements themselves as the representations are in general
no longer ordinary numbers, but instead are matrices of dimension
$> 1$, and \refeq{eq:AbelianGroupRep} no longer applies. 
For $SU(2)$, we choose to label the representations by
the total spin $s$, being related to the eigenvalue of
of the spin operator, $\vektor{S}^2 = s(s+1)$. Assuming that
all of the required operations can be formulated in terms of
manipulations of these representations, we have a formulation
that is \textit{manifestly} $SU(2)$ invariant; 
the rotational invariance is preserved at all steps and at no time
in the calculation is it necessary to choose an axis of 
quantization \cite{NonAbelian}.
This supersedes the earlier approach based on the Clebsch-Gordan
coefficients \cite{OldSU2}.
The non-Abelian formulation is an important optimization, because it increases
the performance of the algorithm by an order of magnitude or 
more \cite{NonAbelian} compared with the Abelian case, 
while enabling more accurate and detailed 
information about the ground-state magnetization.
The basic ingredient that enables this rotationally invariant
construction is the
Wigner-Eckart theorem \cite{WignerSymmetry}, which we can state as:
When written in an angular momentum basis, 
each matrix element of an irreducible tensor operator is a product 
of two factors, 
a purely angular momentum dependent factor (the ``Clebsch-Gordan 
coefficient'') and a factor
that is independent of the projection quantum numbers 
(the ``reduced matrix element''). We formulate the algorithm in such a way
that we store and manipulate only the reduced matrix elements, factorizing
out completely the Clebsch-Gordan coefficients.
The efficiency improvement resulting from the non-Abelian formulation
is that the matrix dimensions $m$ and $M$ now refer to the number of
irreducible representations in the basis, which is typically much
smaller than the total degree of the representation. For a scalar
state, this equivalence is precise: a single representation of degree $N$
in the non-Abelian approach results in $N$ degenerate eigenstates
when the symmetry is not used, with a corresponding improvement in efficiency.
We do not give here a full introduction
to the theory of quantum angular momentum, rather we present,
in the style of a reference, the important formulas required to
manipulate MPS wavefunctions and operators. For a comprehensive
introduction see for example references  \cite{Biedenharn,Varshalovich}.

Using the normalization convention of Biedenharn \cite{Biedenharn}, we define
the matrix elements of a tensor operator $\vektor{T}^{[k]}$ 
transforming as a rank $k$ tensor under $SU(2)$ rotations, as
\begin{equation}
\bigbraket{j'm'}{T^{[k]}_M}{jm}
= \rbigbraket{j'}{\vektor{T}^{[k]}}{j} \; \qcg{j}{k}{j'}{m}{M}{m'} \; ,
\end{equation}
where $C^{\cdots}_{\cdots}$ is the Clebsch-Gordan (CG) coefficient,
$j'j$ label the representation of $SU(2)$, and $m = -j, -j+1,\ldots,j$ and
$m' = -j', -j'+1,\ldots,j'$ label the projections of the total spin
onto the $z$-axis.
Using the orthogonality of the Clebsch-Gordan coefficients, this defines the
reduced matrix elements,
\begin{equation}
\rbigbraket{j'}{\vektor{T}^{[k]}}{j}
= \sum_{mM} \qcg{j}{k}{j'}{m}{M}{m'} \bigbraket{j'm'}{T^{[k]}_M}{jm} \; ,
\end{equation}
where $m'$ is arbitrary. 
Note that this normalization is \textit{not} the same as that used by
Varshalovich \textit{et.~al} \cite{Varshalovich}, whom instead use an additional
factor $\sqrt{2k+1}$ in the reduced matrix elements. This 
is a tradeoff; some formulas simplify slightly with this normalization,
but the normalization used here has the advantage that
the reduced matrix elements of scalar operators (with $k=0$) coincide
with the actual matrix elements as all of the relevant Clebsch-Gordan
coefficients are equal to unity. Given the definition of the reduced matrix
elements, we formulate the remaining formulas without further reference
to the axis of quantization, except as an intermediate step to relate
the reduced matrix elements prior to factorizing out the 
Clebsch-Gordan coefficients.

The coupling of two operators is just as for the coupling of
ordinary spins;
\begin{equation}
\left[ \itensor{S}{k_1} \times \itensor{T}{k_2} \right]^{[k]} \; ,
\end{equation}
which denotes the set of operators with components
\begin{equation}
\left[ \itensor{S}{k_1} \times \itensor{T}{k_2} \right]^{[k]}_{\mu}
=
\sum_{\mu_1 \mu_2} \qcg{k_1}{k_2}{k}{\mu_1}{\mu_2}{\mu}
\itensorcomp{S}{k_1}{\mu_1} \itensorcomp{T}{k_2}{\mu_2} \; .
\end{equation}
Applying the Wigner-Eckart theorem gives, after a few lines of algebra,
\begin{equation}
\begin{array}{c}
\rbigbraket{j'}{\left[ \itensor{S}{k_1} \times 
\itensor{T}{k_2} \right]^{[k]}}{j}
\\
= (-1)^{j+j'+k} \sum_{j''} \sqrt{(2j''+1)(2k+1)} 
\qsixj{j'}{k_1}{j''}{k_2}{j}{k}
\\ \times
\rbigbraket{j'}{\itensor{S}{k_1}}{j''}
\rbigbraket{j''}{\itensor{T}{k_2}}{j} \; ,
\end{array}
\label{eq:IrredProduct}
\end{equation}
where $\{\cdots\}$ denotes the $6j$ coefficient \cite{Biedenharn,Varshalovich}.

A special case of the coupling law \refeq{eq:IrredProduct} that we
will need is when the operators act on different spaces, such that they have
a tensor product form
\begin{equation}
\begin{array}{rcl}
\itensorcomp{S}{k_1}{\mu_1} & = & 
\itensorcomp{T}{k_1}{\mu_1}(1) \otimes I(2) \; , \\
\itensorcomp{T}{k_2}{\mu_2} & = & 
I(1) \otimes \itensorcomp{T}{k_2}{\mu_2}(2) \; .
\end{array}
\end{equation}
Here $I(i)$ denotes the identity operator and 
$\itensor{T}{k_i}(i)$ is an irreducible
tensor operator with respect to the angular momentum $\vektor{J}(i)$
of part $i$ of a two-part physical system ($i = 1,2$). 
The total angular momentum
of the system is $\vektor{J} = \vektor{J}(1) + \vektor{J}(2)$. 
In this case, we write the coupling
as $\itensorcoupling{\itensor{S}{k_1}}{\itensor{T}{k_2}}{k}$
$\equiv$ $\itensoroutercoupling{\itensor{T}{k_1}(1)}{\itensor{T}{k_2}(2)}{k}$.
Repeated application of the Wigner-Eckart theorem to these 
tensor operators gives,
after some algebra,
\begin{equation}
\begin{array}{l}
\rbigbraket{j' \, (j_1'j_2'\alpha_1'\alpha_2')}{
\itensoroutercoupling{\itensor{T}{k_1}(1)}{\itensor{T}{k_2}(2)}{k}}
        {j \, (j_1j_2\alpha_1\alpha_2)} \vspace{0.25cm} \\
        = \qninejsq{j_1}{j_2}{j}{k_1}{k_2}{k}{j'_1}{j'_2}{j'}
        \rbigbraket{j'_1 \, (\alpha'_1)}{\itensor{T}{k_1}(1)}{j_1 \, (\alpha_1)}
        \rbigbraket{j'_2 \, (\alpha'_2)}{\itensor{T}{k_2}(2)}{j_2 \, (\alpha_2)} \; ,
\end{array}
\label{eq:TensorProductCoupling}
\end{equation}
where
\begin{equation}
\qninejsq{j_1}{j_2}{j}{j_1}{k_2}{k}{j'_1}{j'_2}{j'} \equiv 
        [(2j'_1+1)(2j'_2+1)(2j+1)(2k+1)]^{\frac{1}{2}} 
        \qninej{j_1}{j_2}{j}{k_1}{k_2}{k}{j'_1}{j'_2}{j'} \; ,
\end{equation}
and the term in curly brackets is the Wigner $9j$ coefficient, which can be defined as
a summation over $6j$ coefficients \cite{Biedenharn},
\begin{equation}
\begin{array}{rcl}
\qninej{j_1}{j_2}{j}{k_1}{k_2}{k}{j'_1}{j'_2}{j'} & \equiv &
        (-1)^{j_1+j_2+j+k_1+k_2+k+j'_1+j'_2+j'}
        {\displaystyle \sum_{j''} } (-1)^{2j''} (2j''+1) \\
        & & \times
        \qsixj{j'}{k_1}{j''}{k_2}{j}{k}
        \qsixj{j'}{j'_2}{j'_1}{j_1}{k_1}{j''}
        \qsixj{j''}{j_1}{j'_2}{j_2}{k_2}{j} \; .
\end{array}
\end{equation}

We can define an operator norm, corresponding to the usual
Frobenius norm, such that
\begin{equation}
||\itensor{X}{k}||^2_{\mbox{\tiny frob}} = \Tr \itensor{X}{k} \cdot 
\itensordag{X}{k} = 
\Tr \itensordag{X}{k} \cdot \itensor{X}{k} \; .
\end{equation}
After some arithmetic, we see that
\begin{equation}
||\itensor{X}{k}||^2_{\mbox{\tiny frob}} =
\sum_{j'j} (2j'+1) |\rbigbraket{j'}{\itensor{T}{k}}{j}|^2
\end{equation}
For the center-matrix formalism, we need the transformation
\begin{equation}
\itensor{A}{s}_{ij} \rightarrow \sum_k \; C_{ik} \; \itensor{A'}{s}_{kj}
\end{equation}
where $k$ is a $d \times m$ dimensional index that encapsulates 
both a $s'$ and a $j'$ 
index\footnote{More precisely, $k$ runs over the Clebsch-Gordan expansion
of $s' \otimes j'$.}
: $k \simeq (s',j')$.
Requiring $\itensor{A'}{s}_{kj}$ to satisfy the right orthogonality constraint,
$A' A'^{\dagger} = 1$, this requires
\begin{equation}
\itensor{A'}{s}_{kj} = \delta_{j'j}\delta_{s's} \quad \left[\mbox{with }
 k \simeq (s',j')\right]
\end{equation}
with
\begin{equation}
C_{ik} = \itensor{A}{s'}_{ij'}
\end{equation}
In the other direction, we need
\begin{equation}
\itensor{A}{s}_{ij} \rightarrow \sum_k \; \itensor{A'}{s}_{ik} \; C_{kj}
\end{equation}
where $k \simeq (s',i')$.
Requiring $\itensor{A'}{s}_{ik}$ to satisfy the left orthogonality constraint,
$A'^{\dagger} A' = 1$, this requires
\begin{equation}
\itensor{A'}{s}_{ik} = \delta_{s's} \delta_{i'i} \sqrt{\frac{2k+1}{2i+1}} 
\quad \left[\mbox{with } k \simeq (s',i')\right]
\end{equation}
and
\begin{equation}
C_{kj} = \itensor{A}{s}_{j'j} \sqrt{\frac{2i+1}{2k+1}}
\end{equation}

The most natural definition for a matrix product operator has two 
lower indices
and three upper,
\begin{equation}
M^{[k]}_{s'i'} {}^{si}
\end{equation}
which transforms as the product of two operators of rank $[k]$, with
 matrix elements
\begin{equation}
\bigbraket{s'q' ; j'm'}{\itensorcomp{M}{k}{r}}{sq ; jm}
= \rbigbraket{s';j'}{\itensor{M}{k}}{s;j} \; \qcg{s}{k}{s'}{q}{r}{q'} \;
\qcg{j}{k}{j'}{m}{r}{m'} \; .
\end{equation}
Note that the product of an operator and a state requires
a contraction of the index $s$, which has the symmetry of
over two \textit{lower} indices, and then shifting the result index $s'$ 
from upper to lower. For $SU(2)$, the required phase factor is
$(-1)^{s+k-s'}$, giving the rule
\begin{equation}
\itensor{B}{s'} = \itensor{(MA)}{s'} 
= \sum_s (-1)^{s+k-s'} \itensor{M}{k}^{s's} \otimes \itensor{A}{s} \; .
\label{eq:OperatorStateProduct}
\end{equation}

The action of a matrix-product operator on another matrix product
operator is
\begin{equation}
\itensor{X}{x} = \itensor{M}{m} \itensor{N}{n} \; ,
\end{equation}
which corresponds to the ordinary (contraction) product in the local
basis and the tensor product in the matrix basis, and therefore results
in the product of a $6j$ and a $9j$ coefficient from equations
\refeq{eq:IrredProduct} and \refeq{eq:TensorProductCoupling} respectively.

For the evaluation of matrix elements, we need the operation
\begin{equation}
F^{'a'}_{i'j'} = \sum_{s',s,i,j,a}
{M}^{s's}_{a'a} {A}^{*s'}_{i'i} {B}^{s}_{j'j} 
{E}^{a}_{ij}
\end{equation}
On expanding out the reduced matrix elements, we see immediately that the
coupling coefficient is
\begin{equation}
\itensor{F'}{a'}_{i'j'} =
\sum_{a,i,j,k,s,s'}
\qninejsq{j}{s}{j'}{a}{k}{a'}{i}{s'}{i'}
\itensor{M}{k}^{s's}_{a'a} \itensor{A}{s'}^*_{i'i} \itensor{B}{s}_{j'j} 
\itensor{F}{a}_{ij}
\end{equation}
Conversely, from the left hand side,
\begin{equation}
E^{'a}_{ij} = \sum_{s',s,i',j',a'}
E^{a'}_{i'j'} M^{s's}_{a'a} A^{*s'}_{i'i} B^{s}_{j'j}
\end{equation}
is
\begin{equation}
\itensor{E'}{a}_{ij} = \sum_{a',i',j',k,s',s}
\frac{2i'+1}{2i+1}
\qninejsq{j}{s}{j'}{a}{k}{a'}{i}{s'}{i'}
\itensor{E}{a'}_{i'j'}
\itensor{M}{k}^{s's}_{a'a} \itensor{A}{s'}^*_{i'i} \itensor{B}{s}_{j'j}
\end{equation}

On interchanging $A \leftrightarrow E$, $B \leftrightarrow F$, 
this becomes the equation
for a direct operator-matrix-product multiply. But using the center-matrix
formalism, we want instead the operation
\begin{equation}
C'_{i'i} = \sum_{j'jk} E^{k}_{i'j'} C_{j'j} F^{k}_{ij} \; ,
\end{equation}
where $C$ and $C'$ transform as scalars, \ie
the quantum numbers impose $i'=i$, $j'=j$.
This is essentially a scalar product $E \cdot F$, and the coupling 
coefficients drop out. 

\section{Conclusions}
\label{sec:Conclusions}

In this paper, we have presented an introduction to the MPS
formulation of the DMRG algorithm for the calculation of ground-
and excited states of one-dimensional lattice Hamiltonians. The MPS
formulation is extremely flexible, allowing the possibility for
algorithms that act on several distinct wavefunctions at once.
The simplest such algorithms are for the fidelity and
expectation values involving unrelated wavefunctions, 
$\braket{\psi}{\phi}$ and $\bigbraket{\psi}{M}{\phi}$, which are
difficult to extract from conventional DMRG. This gives
access to new tools for the analysis of quantum phase transitions,
by measuring the scaling function and exponents
for the fidelity between ground-states
as a function of the interaction strength. In addition, the MPS 
formulation allow optimized versions of algorithms for dynamical
correlations \cite{Weichselbaum,Andreas} and time 
evolution \cite{RipollTime},
which remains a fertile area for continued algorithmic improvements.

Finally, we note that in the simulation of finite temperature states
via a density operator or purification \cite{VerstraeteDensityOperator,
WhiteDensityOperator} in the absence of dissipative terms that
mix the particle numbers between the real and auxiliary systems,
the symmetries of the system are \textit{doubled}, such that the
symmetries of the Hamiltonian are preserved by the real and
auxiliary parts independently. For simulations in a canonical ensemble,
this leads to a significant efficiency improvement that, as far as
we know, has not yet been taken into consideration.

\ack
Thanks to Ulrich Schollw\"ock and Thomas Barthel for many
stimulating conversations. While preparing this manuscript, we learned that
a rotationally invariant formulation using the Clebsch-Gordan
coefficients \cite{OldSU2} has been applied to the TEBD algorithm
for infinite systems \cite{VidalNew}.


\begin{thebibliography}{99}

\bibitem{White}S. R. White, Phys. Rev. Lett. {\bf 69}, 2863 (1992);
Phys. Rev. B {\bf 48}, 345 (1993).

\bibitem{Klumper}A. Kl\"umper, A. Schadschneider and J. Zittartz, 
J. Phys. A \textbf{24}, L955 (1991).

\bibitem{Fannes}M. Fannes, B. Nachtergaele, and R. F. Werner,
Commun. Math. Phys. \textbf{144}, 443 (1992).

\bibitem{Klumper2}A. Kl\"umper, A. Schadschneider and J. Zittartz,
Z. Phys. B \textbf{87}, 281 (1992).

\bibitem{Derria}B Derrida, M R Evans, V Hakim and V Pasquier,
J. Phys. A \textbf{26}, 1493 (1993).

\bibitem{rommer} S. \"Ostlund and S. Rommer, Phys. Rev. Lett.
{\bf 75}, 3537 (1995).

\bibitem{Pechel}M. C. Chung and I. Peschel,
Phys. Rev. B \textbf{64}, 064412 (2001).

\bibitem{Entropy}G. Vidal, J. I. Latorre, E. Rico, and A. Kitaev,
Phys. Rev. Lett. \textbf{90}, 227902 (2003).

\bibitem{SchuchEntropy}N. Schuch, M. M. Wolf, F. Verstraete and
J. I. Cirac, \textit{preprint} arXiv:0705.0292.

\bibitem{LogScaling}V. E. Korepin, Phys. Rev. Lett. \textbf{92}, 096402 (2004).

\bibitem{VerstraetePBC}F. Verstraete, D. Porras, and J. I. Cirac,
Phys. Rev. Lett.\textbf{93}, 227205 (2004).

\bibitem{VerstraeteDensityOperator}F. Verstraete, J. J. Garc\'ia-Ripoll
and J. I. Cirac,
Phys. Rev. Lett. \textbf{93}, 207204 (2004).

\bibitem{Zwolak}Michael Zwolak and Guifr\'e Vidal,
Phys. Rev. Lett. \textbf{93}, 207205 (2004).

\bibitem{WhiteDensityOperator}A. E. Feiguin and Steven R. White,
Phys. Rev. B \textbf{72}, 220401 (2005).

\bibitem{UliTime}A. J. Daley, C. Kollath, U. Schollw\"ock and G. Vidal,
J. Stat. Mech.: Theor. Exp. P04005 (2004).

\bibitem{WhiteTime}Steven R. White and Adrian E. Feiguin,
Phys. Rev. Lett. \textbf{93}, 076401 (2004).

\bibitem{RipollTime}Juan Jos\'e Garc\'ia-Ripoll, 
New J. Phys. \textbf{8}, 305 (2006).

\bibitem{Zanardi}P. Zanardi and N. Paunkovi\'c, Phys. Rev. E \textbf{74},
031123 (2006).

\bibitem{QuantumPhaseMPS}M. Cizzini, R. Ionicioiu and P. Zanardi,
\textit{preprint} cond-mat/0611727.

\bibitem{Huan}Huan-Qiang Zhou and John Paul Barjaktarevic,
\textit{preprint} cond-mat/0701608.

\bibitem{Nielsen}\book{Michael A. Nielsen and Isaac L. Chuang}
{Quantum Computation and Quantum Information}
{Cambridge University Press, 2000}

\bibitem{Legeza}\"O. Legeza and J. S\'olyom, 
Phys. Rev. Lett. \textbf{96}, 116401 (2006) .

\bibitem{QuanLoschmidt}H.T. Quan, Z. Song, X.F. Liu, P. Zanardi 
and C.P. Sun, Phys. Rev. Lett. \textbf{96}, 140604 (2006).

\bibitem{AndreasThesis}Andreas Freidrich, \textit{PhD Thesis}, 
RWTH-Aachen, 2006.

\bibitem{LegezaBoundary}\"Ors Legeza, Florian Gebhard, and J\"org Rissler,
Phys. Rev. B \textbf{74}, 195112 (2006).

\bibitem{VidalThermo}G. Vidal, Phys. Rev. Lett. \textbf{98}, 070201 (2007).

\bibitem{Higham}\book{Nicholas J. Higham}
{Accuracy and Stability of Numerical Algorithms}
{Society for Industrial and Applied Mathematics, Philadelphia, 2002}

\bibitem{AKLT}I. Affleck, T. Kennedy, E. H. Lieb, and H. Tasaki,
Phys. Rev. Lett. \textbf{59}, 799 (1987).

\bibitem{UliReview}U. Schollw\"ock, Rev. Mod. Phys. \textbf{77}, 259 (2005).

\bibitem{MyThesis}I. P. McCulloch, \textit{PhD Thesis},
Australian National University, 2001.

\bibitem{Weichselbaum}F. Verstraete, A. Weichselbaum, U. Schollw\"ock, 
J. I. Cirac, and Jan von Delft,
\textit{preprint} cond-mat/0504305.

\bibitem{Andreas}A. Friedrich, A. K. Kolezhuk, I. P. McCulloch,
and U. Schollw\"ock, Phys. Rev. B \textbf{75}, 094414 (2007).

\bibitem{Next}I. P. McCulloch, \textit{in preparation}

\bibitem{WhiteSingle}S. R. White, 
Phys. Rev. B \textbf{72}, 180403(R) (2005).

\bibitem{SierraVariational}J. Dukelsky, M. A. Mart\'in-Delgado,
T. Nishino and G. Sierra, Europhys. Lett. \textbf{43}, 457 (1998).

\bibitem{NonAbelian}I. P. McCulloch and M. Gul\'acsi, Europhys. Lett.
\textbf{57}, 852 (2002).

\bibitem{Translation}D. Porras, F. Verstraete, and J. I. Cirac,
Phys. Rev. B \textbf{73}, 014410 (2006).





\bibitem{CorrectionVector}T. D. K\"uhner and S. R. White,
Phys. Rev. B \textbf{60}, (1999).

\bibitem{DDMRG}E. Jeckelmann, Phys. Rev. B \textbf{66}, 045114 (2002).

\bibitem{OldSU2}I. P. McCulloch and M. Gul\'acsi,
Phil. Mag. Lett. \textbf{81}, 447 (2001); 
I. P. McCulloch and M. Gul\'acsi,
Aust. J. Phys. \textbf{53}, 597 (2000).

\bibitem{WignerSymmetry}\book{E. P. Wigner}
{Group Theory and Its Applications to the Quantum Mechanics of Atomic Spectra}
{Academic Press, New York, 1959}

\bibitem{Biedenharn}\book{L. C. Biedenharn and J. D. Louck}
        {Angular Momentum in Quantum Physics}
        {Addison-Wesley, Massachusetts, 1981}

\bibitem{Varshalovich}\book
{D. A. Varshalovich, A. N. Moskalev and V. K. Khersonskii}
{Quantum Theory of Angular Momentum}
{World Scientific, Singapore, 1988}

\bibitem{VidalNew}S. Singh, H.-Q. Zhou, and G. Vidal,
\textit{preprint} cond-mat/0701427.

\end{thebibliography}
\end{document}